# Perceived Acceptability of Sensory Augmentation

A Scenario-Based Survey of Tactile Stimuli and Body Locations in a Museum Context

Michel Witter

Centre of Applied Research for Art, Design and Technology, Avans University of Applied Sciences, Breda, Netherlands, mml.witter@avans.nl

Alwin de Rooij

Department of Communication and Cognition, Tilburg School of Humanities and Digital Sciences, Tilburg University, Tilburg, Netherlands, alwinderooij@tilburguniversity.edu

Michel van Dartel

Centre of Applied Research for Art, Design and Technology, Avans University of Applied Sciences, Breda, Netherlands, mf.vandartel@avans.nl

Emiel Krahmer

Department of Communication and Cognition, Tilburg School of Humanities and Digital Sciences, Tilburg University, Tilburg, Netherlands, e.j.krahmer@tilburguniversity.edu

Sensory augmentation (SA) can extend how people perceive and engage with public environments through the body, offering opportunities to support shared multimodal experiences between deaf and hearing visitors. Designing such systems requires understanding which combinations of tactile stimuli and body locations users perceive as acceptable. We examined the perceived acceptability of fictitious yet feasible SA configurations in a museum use-case. Using a modified technology acceptance model, we surveyed 307 deaf and hearing participants about tactile stimulus types and body locations. The results showed significant effects of stimulus type and body location on acceptance, but no interactions with hearing status. Participants generally preferred temperature-based stimuli, subtle movement, texture, and tickling over vibration, and distal over proximal body locations. These findings suggest that tactile SA design should be guided primarily by embodied and contextual factors rather than hearing status, informing future systems for shared multimodal experience in public cultural settings.

## 1 INTRODUCTION

Interaction between deaf and hearing people is often limited by differences in sensory perception, restricting shared experiences such as museum visits. Sensory Augmentation can address this issue by extending perception beyond the senses available to use naturally [41].

Sensory augmentation technology refers to systems that convey information from novel sensory domains through existing modalities [58]. Examples include the Enactive Torch [17], North Sense [52], and VEST [15]. Despite their potential, such technologies are not widely adopted, largely due to insufficiently inclusive design, resulting in usability and acceptance issues [27].

Developing acceptable SA technology requires extensive prototyping and testing [39, 27]. Given the vast design space of possible combinations of stimuli and body location for sensory augmentation, scoping suitable combinations is required for development of such technology. Tactile stimuli are commonly used in sensory augmentation because they can convey information without interfering with vision or hearing and can be integrated into bodily interactions [2, 27, 23]. In our view, sensory augmentation is both a technological and experiential phenomenon. Sensory augmentation (SA) devices translate environmental signals into perceivable stimuli, and perception emerges through the coupling of the body, technology, and environment. Therefore, we use "SA" rather than "SA technology", unless we are referring to specific technology or devices.

Recent TEI work has explored forms of bodily augmentation related to how technology extends the way people sense and act in everyday environments. For example, Wi-Fi Twinge [47] makes ambient Wi-Fi signals perceivable through electrical muscle stimulation, and PneuMa [43] explores pneumatic extensions that support everyday movement. These projects demonstrate how augmentation can be understood as a coupling of the body, technology, and environment. They also highlight a critical design challenge for SA: bodily augmentation depends not only on the information conveyed but also on how, where, and with what tactile quality the body experiences it.

This study investigates whether the acceptability of SA using tactile stimuli differs between deaf and hearing participants. We address the following research question:

RQ: What SA using tactile stimuli are perceived acceptable for both deaf and hearing people?

The next section outlines the theoretical background and presents the design of a survey conducted to assess acceptability.

## 2 THEORETICAL BACKGROUND

### 2.1 Supporting Shared Museum Experiences

Museum visitors generally enjoy engaging experiences that foster meaningful social connections [11, 57]. Museums are increasingly exploring multisensory and inclusive practices. Examples include the Van Abbemuseum's accessibility initiatives [53] and broader curatorial experimentation [17]. However, interaction between deaf and hearing people is often limited by differences in communication and perception, known as the sensory gap [23, 57]. This gap extends beyond language and cannot be fully resolved through shared communication systems alone. Current assistive technologies, such as those enabling text-to-speech or shared sign language, support communication but do not address differences in sensory experience.

SA technology offers an alternative approach by translating imperceptible environmental signals into perceivable stimuli, such as tactile feedback [44]. Examples of SA devices include the feelSpace belt [27], Harbisson’s color-perception device [30], and "X-ray vision" systems [1, 40]. Despite its potential, SA is limited in real-world adoption due to its development in controlled settings, which results in acceptance challenges [12, 28, 30].

Therefore, this study explores what SA, using tactile stimuli, are perceived acceptable for both deaf and hearing users in a shared museum context.

### 2.2 Measuring Perceived Acceptability

Technology acceptance models like TAM [13] and UTAUT [55] are primarily designed for task-oriented systems and are less suited to evaluate acceptability of sensory technologies. Although no acceptance model exists specifically for SA technology, the model introduced by Spagnolli et al. [50] addresses wearable systems that augment perception, operate continuously, and integrate into everyday life. It covers factors such as usefulness, meaningfulness, comfort, and attitude towards technology. Unlike other TAM-based models, the Spagnolli model also addresses anticipated comfort, social acceptability, and perceived intrusion, making it suitable for evaluating acceptance of hypothetical SA technologies.

### 2.3 Tactile Actuators and Their Body Locations

Tactile stimuli like vibration are widely used in SA [15]. They are produced by actuators such as vibration motors, thermoelectric modules [44], and solenoids, which generate stimuli detected by mechanoreceptors in the skin. These stimuli can evoke sensations such as warmth, stroking, buzzing, a grainy texture, and light touch (see Table 1).

Table 1: Tactile stimuli and corresponding sensations

| Category | Example Sensations | Notes |
|---|---|---|
| Vibration | buzzing, humming, pulsing | Very common in haptics |
| Pressure | light touch, deep pressure, poking, tapping | Sense of contact or presence |
| Movement | brushing, stroking, gliding | Directional cues |
| Texture | smooth, rough, grainy, bumpy | Harder to simulate |
| Temperature | warm, cool | Often for affective feedback |
| Pain/discomfort | prickling, sharp, stinging, burning | Usually avoided, unless deliberately studied |
| Tingling | static electricity, pins-and-needles | Electrical stimulation or nerve activation |
| Tickling/Itchiness | ticklish, itchy | Less common in controlled feedback |
| Skin stretch and squeeze | gentle, gradual pulling or tension on the skin | Convey tension or proximity |

Current actuator technologies include soft, inflatable, and electromagnetic designs (e.g., [45, 10]). These designs are well validated in laboratory settings, but they have been studied less in real-world use. Successful deployment depends not only on technical performance, such as perceptual discrimination [37], but also on user integration (e.g., [5, 36]), usability, and acceptance (e.g., [40]).

The acceptability of tactile stimulation depends on context and individual preferences. However, existing studies often rely on small or heterogeneous samples [24, 32, 9], which limits generalization. This is partly due to that few SA devices are readily available. Also, existing examples such as waist-worn navigation belts [26, 34] and the Enactive Torch [17] demonstrate the feasibility of tactile stimulation but cover only a limited range of stimuli.

### 2.4 Hearing Status

Hearing ability exists on a continuum ranging from mild hearing loss to profound deafness. However, audiological classifications alone do not fully capture the diversity of communication practices and identities within deaf and hard-of-hearing populations. In Deaf Studies, a distinction is commonly made between "deaf" (lowercase), which refers to hearing loss as an audiological condition, and "Deaf" (capitalized), which refers to individuals who identify with the Deaf community as a cultural and linguistic group (e.g., [49, 8]).

In our study, we investigate the possible difference in acceptability of the technology between deaf and hearing people. We use the term “hearing status” for the demographical information provided by participants regarding their hearing ability. In the process of identifying hearing status, we use the term “hard-of-hearing” to refer to people with mild to severe hearing loss who still have some functioning hearing, whereas “deaf” refers to people with severe of profound hearing loss.

### 2.5 The Present Study

Research into somatosensory perception and SA indicates that preferences for tactile stimuli depend on both the type of stimulus and the body location at which it is applied, as well as relations to bodily sensitivity [27], the interpretability of signals [2], and social acceptability [38]. A substantial body of work has demonstrated that more peripheral parts of the body are preferred for providing tactile feedback when using similar sensory devices. In addition to body location, different types of tactile stimuli (e.g., vibration, temperature, and pressure) may be experienced and interpreted differently. This could potentially influence their acceptability for use. For instance, apart from the tongue, lips and the face, peripheral body parts have a higher tactile resolution than parts of the central body [2, 23]. Dim and Ren [14] note that fingers, wrists, and feet are among the areas with the highest perceptibility and user preference. Studies on tactile discrimination in the hands [51] and vibrotactile feedback in the feet and ankles [31] also support that these areas are among the most suitable areas for tactile stimulation by SA technology.

Given SA’s potential to bridge the sensory gap described earlier, it is surprising that differences in such preferences between deaf and hearing populations have yet to be researched. This raises the following subquestions:

**SQ1**: How does acceptability of body locations for tactile stimuli differ according to hearing status?

**SQ2**: How does acceptability of tactile stimulus types differ according to hearing status?

Although deaf individuals often rely more on visual cues and use forms of communication that are more grounded in bodily expression than those of hearing individuals, there is currently no established theory connecting these differences to tactile preferences. This raises a third subquestion:

**SQ3**: Which specific body locations are preferred or non-preferred according to hearing status?

Given that development of deployable SA technologies is complex and time-consuming, to address these subquestions, we scope the range of possible combinations of tactile stimuli and body locations by understanding which artificial tactile stimuli are preferred, on which body locations and which differences arise in those preferences between deaf and hearing people.

## 3 METHOD

### 3.1 Study Design

Our study design is based on survey-based studies regarding sensor-based technologies that make use of fictitious use-scenarios, like Heerink et al. [20] and Spagnolli et al. [50], or minimal-exposure use-scenarios, like Kilian et al. [25]. It also builds on the predictive strength of technology acceptance models for wearable sensing devices, as Okpala et al. [33] showed that beliefs about usefulness, effort, and context are sufficient predictors of adoption.

### 3.2 Participants and Sampling

Participants were recruited via Prolific using purposive sampling. Eligible participants were aged 18 or older, fluent in English, deaf or hearing, and had an undergraduate degree or higher. Participants were required to have no additional disabilities. A prescreening question was used to determine hearing status, resulting in the following three participant groups: hearing, hard of hearing, or deaf. An additional hearing sample was recruited using the same criteria. Participants received financial compensation. The survey was distributed to 190 hearing, 141 hard-of-hearing, and 51 deaf participants, yielding 164, 135, and 8 responses, respectively. After data cleaning, the sample comprised 307 participants. Sample characteristics are reported in Table A1 in Appendix A.

### 3.3 Materials and Measurements

In this study, participants were asked to rate the acceptability of a presented use scenario, followed by a technology acceptability questionnaire, an inquiry into preferred sensation types and two questions on participants' preference and non-preference of specific body parts for tactile stimuli.

#### *3.3.1Use scenario.*

Participants are asked to rate acceptability of a SA system in a museum context, presented in the following hypothetical use scenario:

"Imagine you enter an art museum. At the front desk, the attendant offers you a device. They say you will wear the device on your body, and it will provide you sensory feedback as you walk through the museum. The device is responding to invisible elements in the museum. Therefore, with the device you can perceive elements that others cannot perceive unless they wear the same device."

#### *3.3.2Measuring acceptability.*

We deployed an adaptation of an existing technology acceptance model used in a study on user acceptance of wearable symbiotic devices [50], based on both TAM and UTAUT models [55] introduced in the theoretical background section. We modified a few statements from the original list to fit the presented technology and context of use. Appendix A lists all adaptations to the original model. Participants were asked to "Please indicate the extent to which you agree with the following statements where 1 = Extremely disagree and 7 = Extremely agree", which was followed by a 7-point Likert scale.

The full 26-item list can be found in Appendix A and covers factors of perceived usefulness, perceived comfort/pleasantness, facilitating conditions, and attitude toward technology, which Spagnolli at al. [50] consider good predictors of acceptability. For instance, it includes the statements:

- "The device would be incompatible with most aspects of my museum visit."
- "The device limits the way in which I like to visit a museum."
- "The device could help in exploring the museum."

#### *3.3.3Preferred sensation types and body locations.*

To understand what combinations of sensation type and body location are acceptable, we used a matrix question where the participant responded to 8 sensation types (see Table 1) and 4 general body locations using a 5-point Likert scale ranging from most preferred (5) to least preferred (1). Harmful or generally unpleasant sensations (i.e., pain, discomfort) are not considered, leaving eight possible sensation types which we have incorporated in the scenarios. We also provided the participant a silhouette picture like Figure 1 though with only 4 labels referring to body locations.

For accessibility reasons we used the term “sensations” to avoid the more technical term stimuli (see survey contents in Appendix A).

*3.3.4Preference and non-preference for specific body locations.*

To understand which body locations are mostly preferred or not preferred for tactile stimuli, we created a question consisting of a list of 21 body locations and a picture showing a human silhouette with labels referring to body locations (see Figure 1). Participants specify a maximum of three preferred body locations. Subsequently, a second matrix question presents the same list except the selected items from the previous question. Here, the participant specifies a maximum of three body locations they do not prefer.

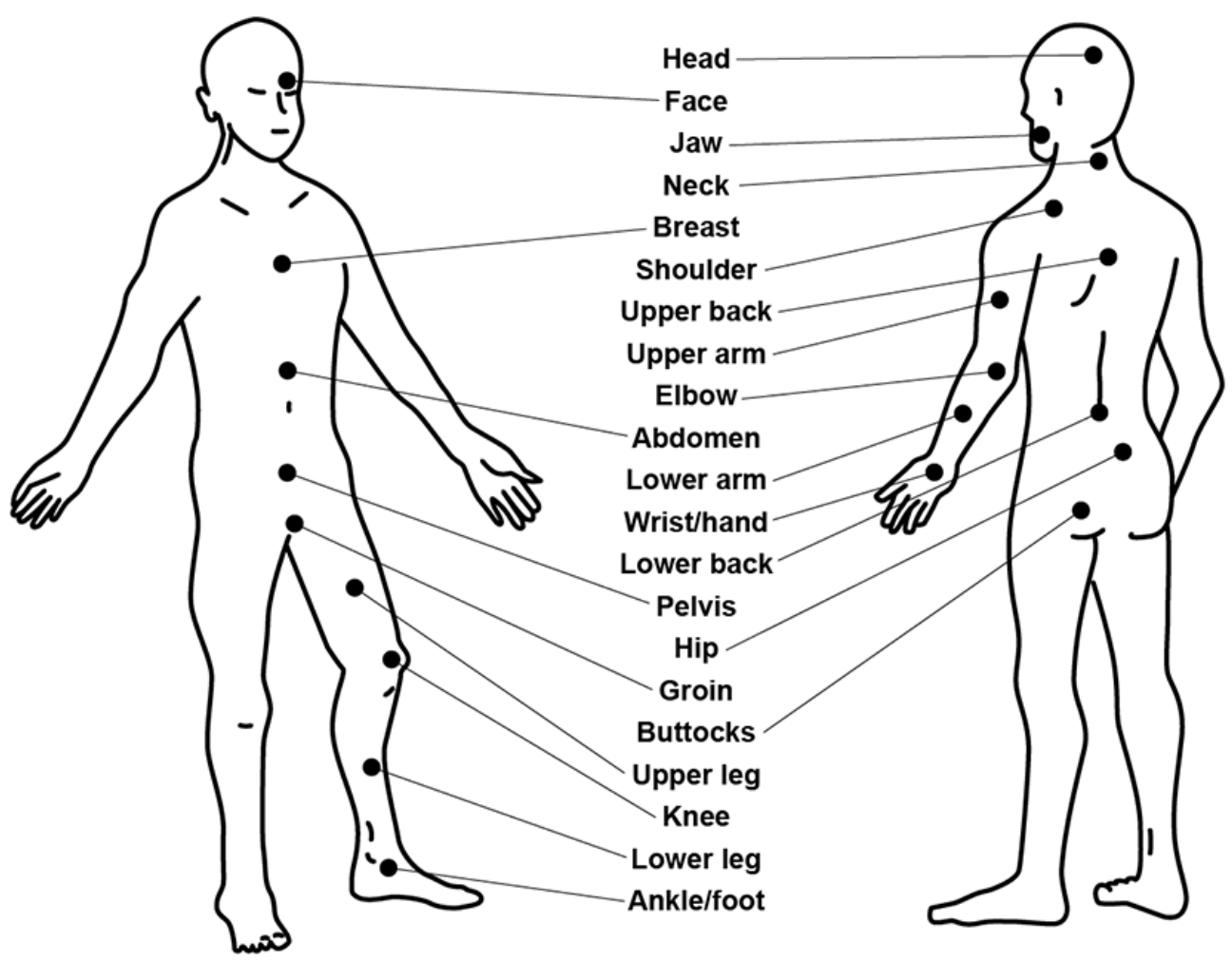

Figure 1: Question type for preferred and non-preferred body locations. Adapted from the Michigan Body Map [7]

## 3.4 Procedure

Participants accessed the survey online via Prolific. The survey opened with an introduction, informing the participant about the study. It also showed a consent form, a demographics section (age, hearing ability, deafness emergence, gender, an acquaintance with museum visits) and an explanation of the presented technology. Only after giving consent, each participant was presented the following questions that we will explain below (see Figure 2, and see Appendix A for the contents of the survey). The study took participants approximately 15 minutes to complete.

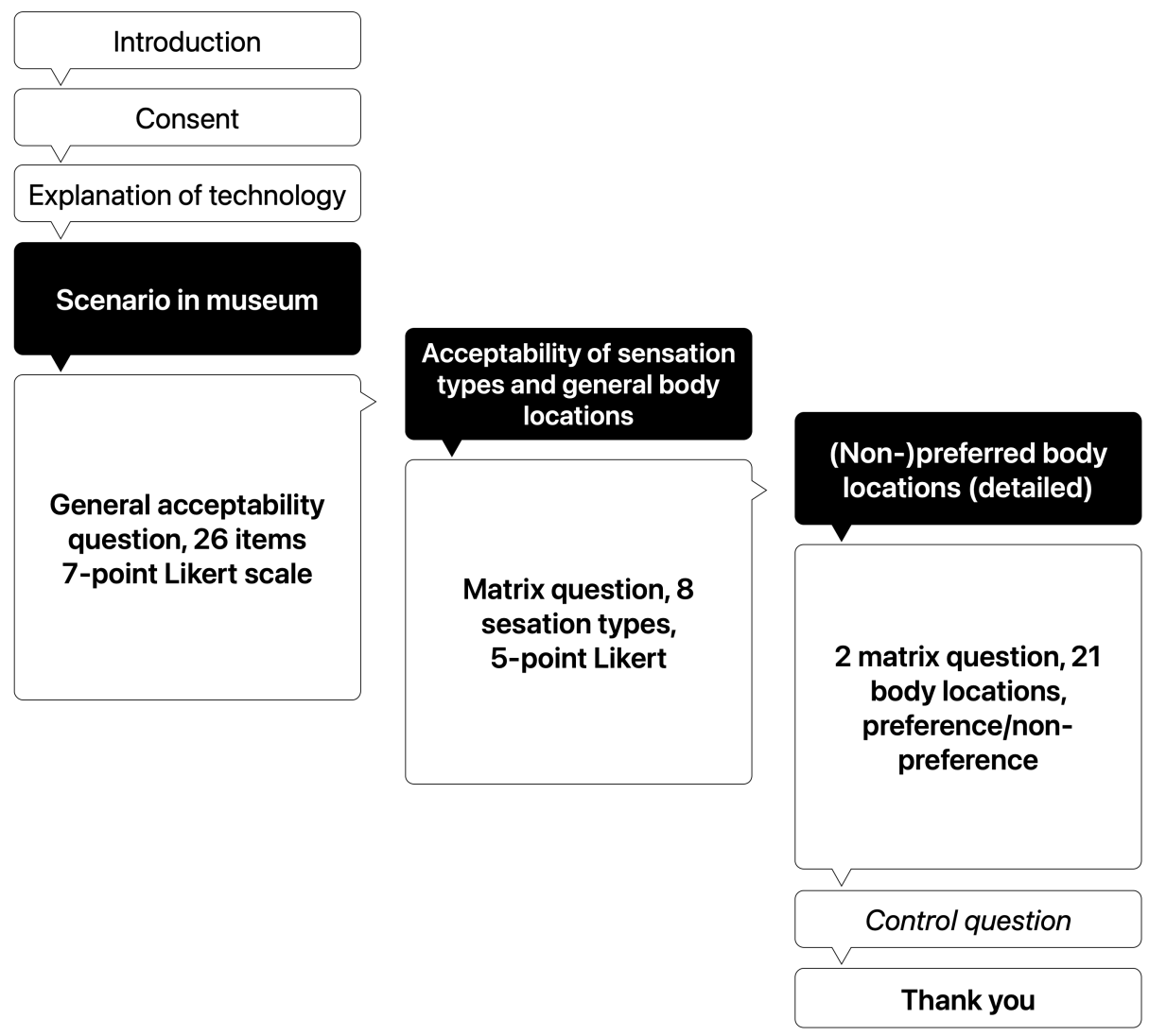

Figure 2: Survey structure

### 3.5 Data Processing and Analysis

In Qualtrics, custom logic was created to divide participants between "Deaf", "Hearing", “I’m not sure” and "Prefer not to answer" groups, based on their responses. Participants in the "Prefer not to answer" and “I’m not sure” groups were excluded from analysis, as well as one participant mentioning age “Under 18”. Also, submissions with incomplete data and that answered wrongly on at least one of both control questions were excluded. This data screening resulted in 307 submissions for analysis.

Microsoft Excel was used for cleaning survey data and determining sample groups for deaf and other hard-of-hearing levels. And we used JASP (version 0.96) for analysis of data from the acceptability survey.

The analysis consisted of the following steps:

1. Differences in acceptability according to hearing status were assessed using Welch’s independent-samples t-test, following reliability analysis and inspection of normality assumptions via Shapiro–Wilk tests and Q–Q plots. Reliability was excellent (Cronbach’s $\alpha = .92$), allowing use of averaged scores. Despite slight deviations from normality (Shapiro–Wilk: $W = 0.98$, $p < .001$), Welch’s t-test was applied due to robustness in large samples.
2. To assess acceptability of sensation types on the body a mixed-design repeated-measures ANOVA was conducted with sensation type as a within-subjects factor and hearing status as a between-subjects factor. Sphericity was assessed using Mauchly’s test, and Huynh–Feldt corrections were applied where this assumption was violated. Normality was evaluated using Q–Q plots. Post-hoc pairwise comparisons were conducted with Holm correction for multiple testing.
3. To assess acceptability of tactile stimuli on general body parts a two-way mixed-design ANOVA was conducted with body part as a within-subjects factor and hearing status as a between-subjects factor. Mauchly’s test indicated a violation of sphericity, and Huynh–Feldt corrections were applied. Holm-corrected pairwise comparisons were used for post-hoc analyses.

4. And, to analyze preferences of detailed body parts we fitted a logistic mixed-effects model (binomial link) with body location as a within-subjects factor and hearing status as a between-subjects factor, including a random intercept for participants. To ensure stable estimation, categories with fewer than ten observations were excluded. Model effects were evaluated using likelihood ratio tests comparing nested models.

## 4 RESULTS

This section reports the main analysis, followed by an additional analysis (see section 5).

### 4.1 General Acceptability According to Hearing Status

We first checked whether there were differences in general acceptability of SA between participants. No significant differences were found between hearing ($M$ = 5.11, $SD$ = 0.82) and hearing-impaired participants ($M$ = 5.18, $SD$ = 0.83), $t(279.30) = 0.66$, $p = .511$, $d = 0.08$. S1 in Appendix B presents further details. These results indicate that hearing status is not associated with general acceptability of SA technologies.

### 4.2 Acceptability of Sensation Types on the Body

Next, we evaluated whether there were overall differences in the acceptability of different sensation types on the body. Figure 3 summarizes the results and inspection of this figure shows that people differently appreciate different sensation types, with Temperature most and Stretch least appreciated. A repeated-measures ANOVA revealed a significant effect of sensation type: $F(6.16, 1878.86) = 55.06$, $p < .001$, $\eta_p^2 = .153$. In line with section 4.1, we found that these appreciations do not depend on hearing status ($F(1, 305) = 0.10$, $p = .748$, $\eta_p^2 < .01$) nor was there an interaction between acceptability and hearing status ($F(6.16, 1878.86) = 0.90$, $p = .500$, $\eta_p^2 < .01$).

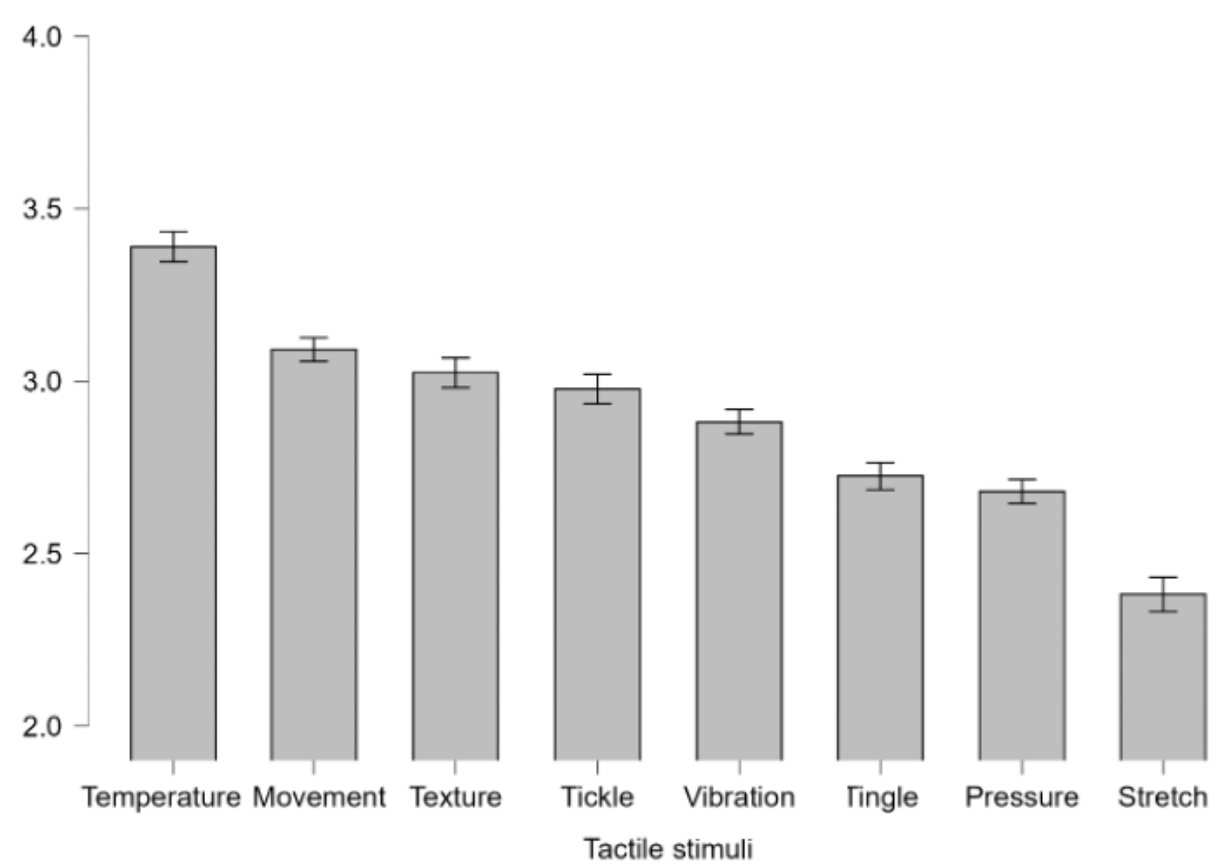

Figure 3: Bar chart displaying the estimated marginal means of the 1-5 acceptability ratings for different technology-actuated sensation types on the body. The error bars represent the standard errors around the means

Holm-corrected post-hoc comparisons indicated that temperature ($M$ = 3.39, $SD$ = 0.95) was the most acceptable sensation, significantly more so than movement, texture, tickling, and vibration (all $p < .001$). Stretching ($M$ = 2.38, $SD$ = 1.14) was the least acceptable sensation, followed by tingling and pressure (all $p < .001$; see S2 in Appendix B).

These findings indicate strong differences across sensation types, independent of hearing status.

### 4.3 Acceptability of Tactile Stimuli on Different Body Parts

Then, we evaluated whether there were differences in acceptability of tactile stimuli on different body parts. Figure 4 summarizes the results and inspection of this figure shows that people differently appreciate different body parts, with arms most and head least appreciated. A repeated-measures ANOVA revealed a significant effect of body part: $F(2.51, 764.17) = 257.10$, $p < .001$, $\eta_p^2 = .46$. Again, in line with section 4.1, we found that these appreciations do not depend on hearing status ($F(1, 305) = 0.10$, $p = .748$, $\eta_p^2 < .01$) nor was there an interaction between acceptability and hearing status ($F(2.51, 764.17) = 2.12$, $p = .108$, $\eta_p^2 = .01$).

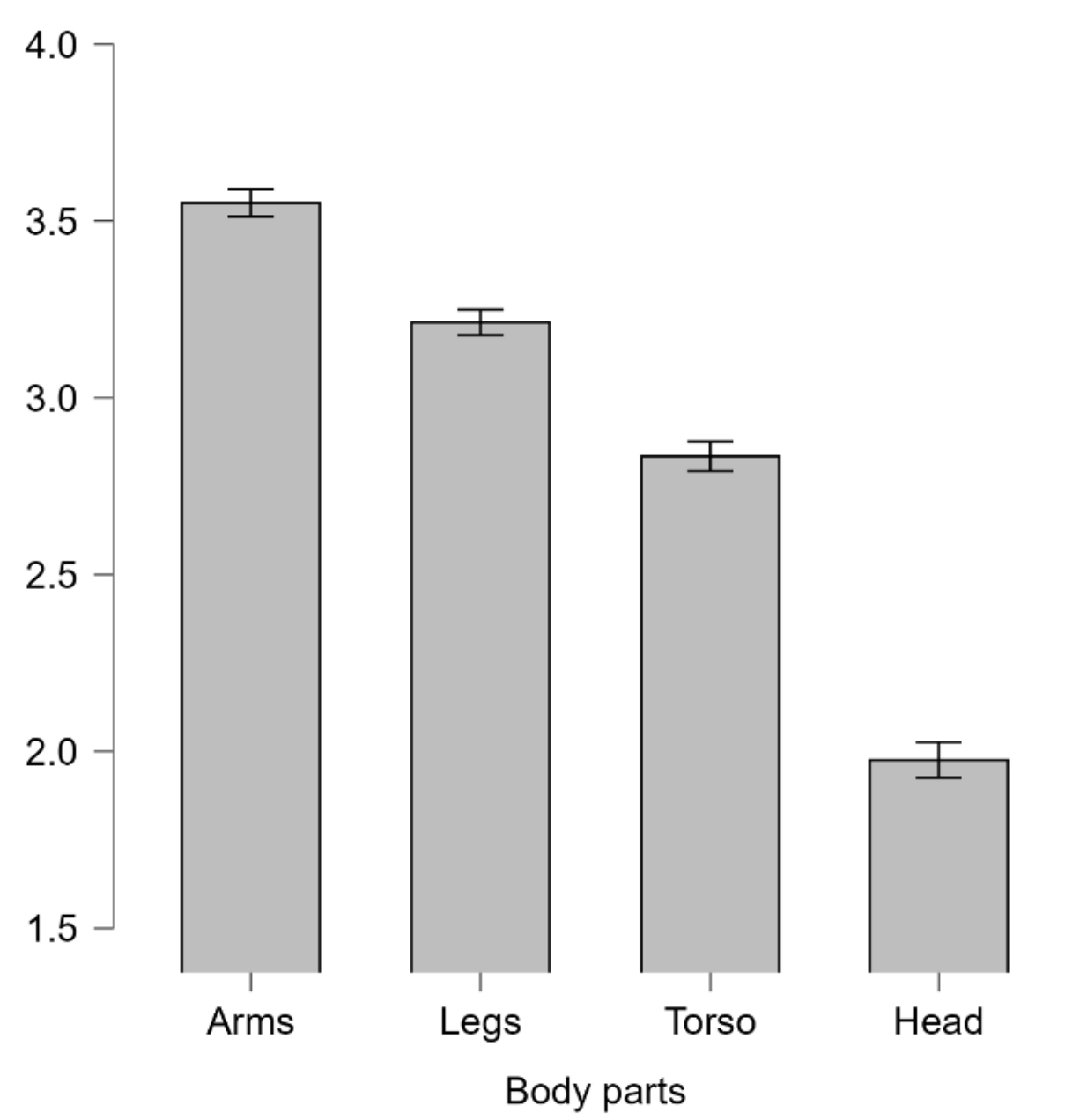

Figure 4: Bar chart displaying the estimated marginal means of the 1-5 acceptability ratings for using technology for tactile stimuli on different body parts. The error bars represent the standard errors around the means

Post-hoc comparisons revealed that the arms ($M = 3.55$, $SD = 1.015$) were the most acceptable body part, followed by the legs the head (all $p < .001$). The head was consistently the least acceptable body part.

Together, these results demonstrate clear differences across body parts with no effect of hearing status.

### 4.4 Preference and Non-preference for Specific Body Locations

And, finally, we evaluated the preference and non-preference of 21 specific body locations (see Table 2). Here, preferences varied significantly across body locations: $\chi^2(16) = 322.04$, $p < .001$. However, neither the interaction with hearing status ($\chi^2(16) = 17.64$, $p = .345$) nor the main effect of hearing ($\chi^2(1) = 0.94$, $p = .333$) was significant. And non-preferences exhibited a comparable trend ($\chi^2(12) = 260.29$, $p < .001$) with no significant interaction or the main effect of hearing status.

Descriptively, preferred locations included the wrist, hand, lower arm, upper arm, and shoulder. Non-preferred locations included the head, face, groin, breast, jaw, buttocks, pelvis, and neck (see Table 2 and S4 in Appendix B for further details).

Table 2: Frequency and percentages of preferred and not preferred body locations for technology-actuated tactile stimuli

| Body location | Preferred | | Not preferred | |
|---|---|---|---|---|
| | n | % | n | % |
| Abdomen | 33 | 3.84 | 33 | 3.68 |
| Ankle/foot | 34 | 3.95 | 18 | 2.01 |
| Breast | 27 | 3.14 | 86 | 9.59 |
| Buttocks | 5 | 0.58 | 63 | 7.02 |
| Elbow | 16 | 1.86 | 5 | 0.56 |
| Face | 17 | 1.98 | 147 | 16.39 |
| Groin | 2 | 0.23 | 117 | 13.04 |
| Head | 30 | 3.49 | 178 | 19.84 |
| Hip | 14 | 1.63 | 11 | 1.23 |
| Jaw | 2 | 0.23 | 67 | 7.47 |
| Knee | 10 | 1.16 | 14 | 1.56 |
| Lower arm | 128 | 14.88 | 3 | 0.33 |
| Lower back | 23 | 2.67 | 10 | 1.11 |
| Lower leg | 48 | 5.58 | 5 | 0.56 |
| Neck | 16 | 1.86 | 59 | 6.58 |
| Pelvis | 4 | 0.47 | 60 | 6.69 |
| Shoulder | 84 | 9.77 | 2 | 0.22 |
| Upper arm | 126 | 14.65 | 4 | 0.45 |
| Upper back | 57 | 6.63 | 5 | 0.56 |
| Upper leg | 25 | 2.91 | 6 | 0.67 |
| Wrist/hand | 159 | 18.49 | 4 | 0.45 |

Note: data are frequencies (n) and percentages (%) rounded to two decimal places for preferred (left) and not preferred (right) body locations.

These findings support our earlier observations that the arms and legs are generally preferred locations for tactile stimuli, while offering more fine-grained insight into which specific locations within these regions are most acceptable, and which areas, particularly on the head, face, and torso, are consistently avoided.

## 5 ADDITIONAL ANALYSES

The preceding analyses compared participants with and without hearing impairments. To examine differences within the hearing-impaired group, analyses were repeated using three levels of self-reported hearing impairment: mild (n = 103), moderate (n = 30), and severe/deaf (n = 8; see Table 1). Due to the small sample sizes of those with severe impairment (n = 5) and deafness (n = 3), these two groups were combined to ensure stable estimates.

### 5.1 General Acceptability Across Degrees of Hearing Impairment

A one-way ANOVA revealed no significant differences in general acceptability across groups ($F(2, 138) = 1.85$, $p = .161$, $\eta_p^2 = 0.03$.

### 5.2 Acceptability of Sensation Types

Acceptability of tactile stimuli across levels of hearing impairment is shown in Figure 5. A repeated measures ANOVA revealed a significant main effect of sensation type ($F(5.87, 810.35) = 13.47$, $p < .001$, $\eta_p^2 = 0.09$), which replicated the earlier pattern. In line with section 4.1, no interaction with degree of impairment was found. However, a small main effect of hearing impairment was observed: $F(2, 138) = 3.53$, $p = .032$, $\eta_p^2 = 0.05$.

Post-hoc comparisons revealed a similar ranking as in the main analysis. Temperature was most acceptable ($M = 3.36$, $SD = 0.94$), though not significantly more so than movement or texture. It was, however, rated significantly higher than tickling and vibration (all $p < .001$). The least acceptable stimuli were stretching, followed by tingling and pressure. See S6 in Appendix B for the complete post-hoc comparison results.

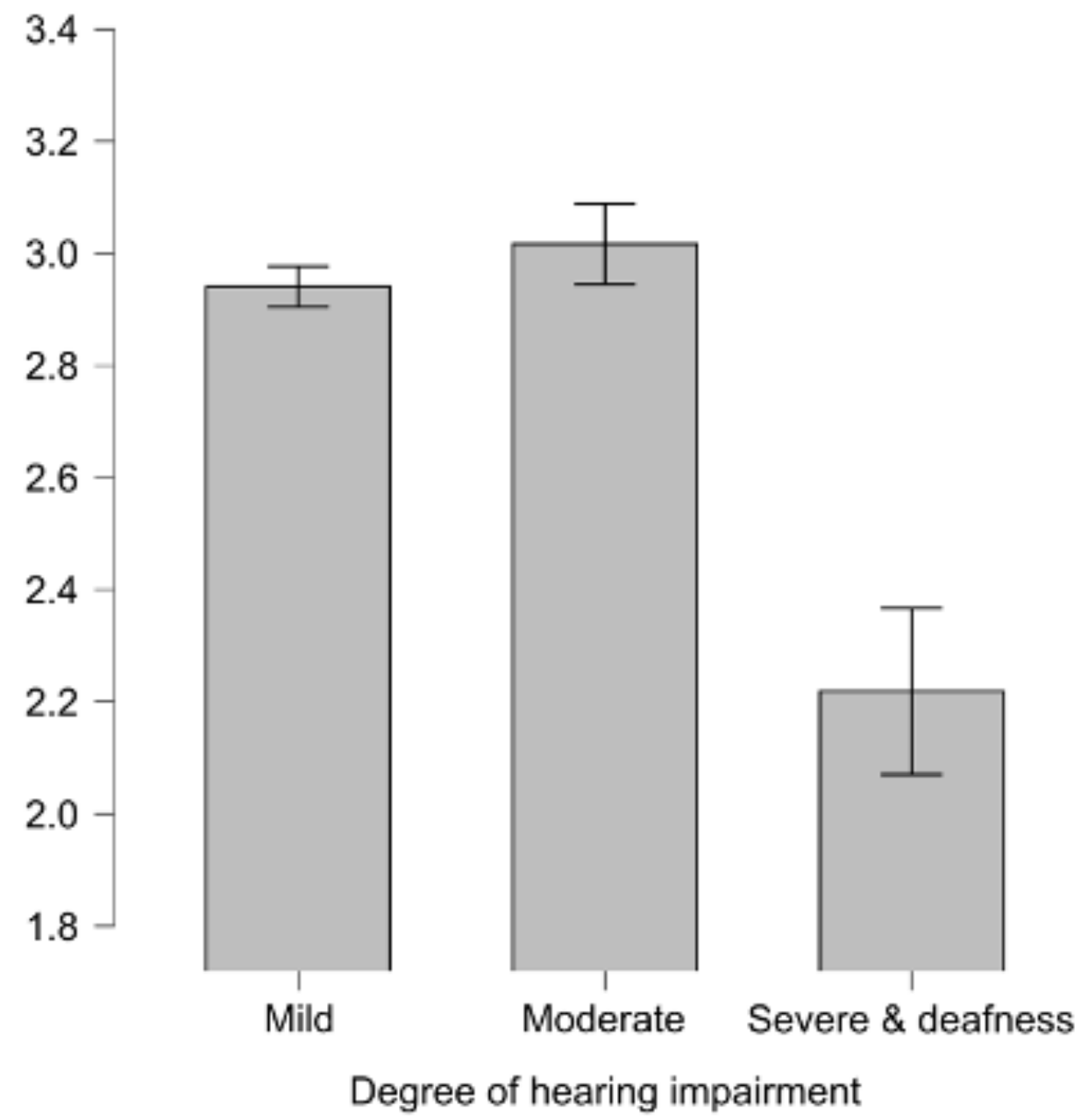

Figure 5: Differences between degrees of hearing impairment regarding the acceptability of tactile stimuli

Further analysis of Holm-corrected comparisons between hearing-impairment groups showed that participants with severe hearing impairment or rated tactile stimuli as less acceptable overall than participants with mild or moderate hearing impairment. However, no significant difference was found between mild and moderate hearing impairment ($t(138) = 0.47$, $p = .637$, $d = 0.07$). See S6 in Appendix B for more details.

These findings largely mirror the earlier pattern across sensation types. They also suggest lower overall acceptability among participants with severe hearing impairment or deafness.

### 5.3 Acceptability Across Body Parts

Next, we evaluated whether there were differences in acceptability of tactile stimuli on different body parts across levels of hearing impairment. A repeated-measures ANOVA revealed a significant main effect of body part ($F(2.54, 350.10) = 39.20$, $p < .001$, $\eta_p^2 = 0.22$), which replicated the earlier pattern (see Figure 4 and Appendix A for the corresponding bar chart). Mauchly's test indicated a violation of sphericity ($W = 0.72$, $p < .001$), so Huynh–Feldt corrections were applied. There was no interaction between body part and degree of hearing impairment ($F(5.07, 350.10) = 0.29$, $p < .001$, $\eta_p^2 < 0.01$). However, a small but significant main effect of hearing impairment was observed ($F(2, 138) = 3.53$, $p = .032$, $\eta_p^2 = 0.05$), which is consistent with the findings reported in Section 5.2.

Holm-corrected post-hoc comparisons confirmed the same pattern of acceptability across body parts within the hearing impairment sample. Participants rated tactile stimuli on the arms as the most acceptable ($M = 3.53$, $SD = 1.03$), followed by the legs ($M$ = 3.18, $SD$ = 0.98), the torso ($M$ = 2.91, $SD$ = 1.05) and head ($M$ = 2.04, $SD$ = 0.89). The arms were rated significantly more acceptable than the legs, torso, and head (all $p \leq .005$). The legs were rated as more acceptable than the torso and head (all $p < .001$. The torso was rated as more acceptable than the head (all $p < .001$; see S7 in Appendix B for the complete post hoc comparisons).

These finding replicate the earlier pattern of body-location preferences while indicating slightly lower overall acceptability among participants with more severe hearing impairments. For this analysis, the same data is used as in the previous analysis, albeit being aggregated differently. We therefore refer to the subsection above, S6 and S7 in Appendix B for details.

### 5.4 Preference and Non-preference for Specific Body Locations Across Degrees of Hearing Impairment

Finally, we also evaluated the preference and non-preference of the specific body locations (as in Table 2) across different levels of hearing impairment. Logistic mixed-effect models were fitted to examine whether the degree of hearing impairment relates to patterns of preferred and non-preferred body locations. Body location (21 categories; see a similar distribution in Table 2) was used as a within-subjects factor, and hearing impairment was used as a between-subjects factor. As in the main analysis, categories with fewer than ten observations were excluded to ensure stable estimates.

For preferred locations, significant variation across body locations was observed ($\chi^2(11) = 138.90$, $p < .001$). However, neither the interaction between body location and hearing status ($\chi^2(22) = 18.54$, $p = .673$) nor the main effect of hearing status ($\chi^2(2) = 0.27$, $p = .874$) was significant.

A similar effect was found for non-preferred locations. After excluding sparse categories (see a similar distribution in Table 2), significant variation across body location was observed ($\chi^2(8) = 83.54$, $p < .001$). There was no significant interaction ($^2(16) = 10.66$, $p = .830$) or main effect of hearing status ($\chi^2(1) = 0.02$, $p = .891$).

Descriptively, preferred locations included the wrist/hand, upper arm, lower arm, and shoulder, while non-preferred locations included the head, face, breast, buttocks, neck, and pelvis. For detailed post-hoc comparisons, see S8 in Appendix B.

These results mirror those of the main analysis, indicating that preferences vary across body locations but are not influenced by hearing status.

## 6 DISCUSSION

### 6.1 Main Findings

Four main findings emerged: two empirical effects (stimulus type and body location), one null effect (no difference between deaf and hearing participants), and one theoretical implication (the limited role of sensory background in shaping technology acceptability).

Regarding SQ1 "*How does acceptability of body locations for tactile stimuli differ according to hearing status?*", no differences were found in acceptability or preferences for tactile stimuli and body locations, between deaf and hearing participants. Although tactile signals may be interpreted differently [3], these differences do not translate into preferred body locations or stimuli for SA technology.

With respect to SQ2 "*How does acceptability of tactile stimulus types differ according to hearing status?*", our findings contrast with studies suggesting deaf individuals attribute greater informational value to tactile cues [4, 18]. Instead, our results suggest that vibrotactile stimuli are less preferred than other tactile stimuli by both deaf and hearing participants.

For SQ3 "*Which specific body locations are preferred or non-preferred according to hearing status?*", our results align with prior work showing that preferred body locations are largely stable across user groups and driven by perceived usefulness and situational context than by hearing status [46, 55]. This is noteworthy since deaf rely more on vision and touch for environmental awareness, navigation and social attunement than hearing people do.

Taken together, these findings suggest that the perceived acceptability of SA is more influenced by embodied design factors than by hearing status. These factors include the tactile quality of the stimulus, the body location at which it is applied, and how well it fits with everyday bodily activity. This interpretation aligns with recent TEI research on bodily augmentation. For example, the Wi-Fi Twinge [47] project demonstrates how imperceptible environmental signals can be made available through wearable stimulation; and the PneuMa project [43] explores bodily extensions that support everyday movement. These projects demonstrate that augmentation occurs through the coupling of the body, technology, and the environment. Our findings extend this line of work by showing that such coupling is also constrained by perceived acceptability. Where and how the body is stimulated matters before a system is deployed.

### 6.2 Limitations

Our study has three limitations, related to hypothetical versus real use, sample imbalance across hearing statuses, and procedural constraints in splitting the sample during analysis.

First, using hypothetical scenarios allows for a systematic comparison of tactile stimuli and body locations, yet it limits ecological validity. Participants evaluated SA configurations without wearing, learning, or socially negotiating how the stimuli would be used. This is particularly relevant for embodied and tactile systems. Sensory feedback wearables may alter how users perceive and experience their own body [54], and the user can acquire affective or interpersonal meaning through actual wearable tactile interactions [42]. Therefore, the effective use of novel sensory feedback requires integration into sensorimotor models over time, making usability, cognitive load, and experiential value difficult to judge from scenarios alone. Consequently, our findings should be interpreted as reflecting initial perceived acceptability rather than acceptance after adaptation. Nevertheless, the absence of differences between deaf and hearing participants is consistent with existing theories on tactile perception (e.g., [19] and [21]) and tactile acuity (e.g., [35] and [6]).

Secondly, sample imbalance limits the reliability of subgroup comparisons. While the overall patterns remained consistent across hearing groups, the small number of participants with severe hearing loss or who were fully deaf introduces uncertainty when interpreting lower acceptability scores. Further research with larger, more balanced samples is necessary for more reliable comparison.

Thirdly, selecting group members from the sample by combining participants in Qualtrics and separating groups post hoc using Prolific IDs requires time and attention. Incorporating a marker question could have simplified group assignment.

## 7 CONCLUSION

This paper contributes to research on tangible and embodied interaction by examining the perceived acceptability of tactile sensory augmentation (SA) in a museum setting. No significant differences were found between deaf and hearing participants across all three sub-questions, challenging the assumption that hearing status should determine design choices for sensation type or body locations. Rather than suggesting that hearing status should determine the design of tactile SA, our findings indicate that the acceptability of SA in shared public environments depends primarily on embodied design factors, such as tactile quality, body location, and the anticipated fit between the body, technology, and context.

In response to our main RQ "*What SA using tactile stimuli are perceived acceptable for both deaf and hearing people?*", we identified a range of acceptable tactile stimuli and body locations, providing empirical guidance for designing sensory augmentation technologies for public, multimodal environments.

## ACKNOWLEDGMENTS

During manuscript preparation, the authors used DeepL and ChatGPT (version 5.2) to assist with language editing and improving clarity. All outputs were reviewed and verified by the authors.

This publication is part of the project "*Gesitueerde ontwerpmethoden voor het ontwerpen van sensory augmentation-toepassingen voor doven*" with project number 023.017.033 of the research program "*Promotiebeurs voor leraren – PvL 2021-1*" which is partly financed by the Dutch Research Council (NWO).